 \documentclass[onecolumn,apj,numberedappendix]{openjournal}

\usepackage{xspace}
\usepackage{hyperref}

\usepackage[normalem]{ulem}
\usepackage{orcidlink}
\usepackage{soul}
\usepackage{etoolbox}

\newcommand\radioastrotools{\texttt{radio-astro-tools}\xspace}
\newcommand\spectralcube{\texttt{spectral-cube}\xspace}
\newcommand{\radiobeam}{\texttt{radio-beam}\xspace}
\newcommand{\pvextractor}{\texttt{pvextractor}\xspace}
\newcommand{\uvcombine}{\texttt{uvcombine}\xspace}
\newcommand{\casaformatsio}{\texttt{casa-formats-io}\xspace}

\newcommand{\astropy}{\texttt{astropy}\xspace}
\newcommand{\dask}{\texttt{dask}\xspace}
\newcommand{\reproject}{\texttt{reproject}\xspace}
\newcommand{\matplotlib}{\texttt{matplotlib}\xspace}
\newcommand{\regions}{\texttt{regions}\xspace}

\begin{document}

\title{radio-astro-tools: linking radio astronomical data to the astronomical Python ecosystem}

\author{Eric W. Koch\orcidlink{0000-0001-9605-780X}}
\affiliation{Dominion Radio Astrophysical Observatory, Herzberg Astronomy \& Astrophysics, National Research Council Canada, P.O. Box 248, Penticton, BC V2A 6J9,
Canada}
\email{koch.eric.w@gmail.com}

\author{Adam Ginsburg\orcidlink{0000-0001-6431-9633}}
\affiliation{Department of Astronomy, University of Florida, PO Box 112055, USA}
\email{adam.g.ginsburg@gmail.com}

\author{Thomas~P.~Robitaille\orcidlink{0000-0002-8642-1329}}
\affiliation{Aperio Software, Insight House, Riverside Business Park, Stoney Common Road, Stansted, UK}

\author{Erik~W.~Rosolowsky\orcidlink{0000-0002-5204-2259}}
\affiliation{Dept. of Physics, University of Alberta, 4-183 CCIS, Edmonton, Alberta, T6G 2E1, Canada}


\author{Alyssa Bulatek\orcidlink{0000-0002-4407-885X}}
\affiliation{Department of Physics and Astronomy, Haverford College, 370 Lancaster Avenue, Haverford, PA 19041, USA}

\author{Savannah Gramze\orcidlink{0000-0002-1313-429X}}
\affiliation{Department of Astronomy, University of Florida, PO Box 112055, USA}

\author{Jonathan~D.~Henshaw\orcidlink{0000-0001-9656-7682}}
\affiliation{Max-Planck-Institut f\"ur Astronomie, K\"onigstuhl~17, D-69117 Heidelberg, Germany}

\author{Derek Homeier}
\affiliation{Aperio Software, Insight House, Riverside Business Park, Stoney Common Road, Stansted, UK}

\author{Preshanth Jagannathan\orcidlink{0000-0002-5825-9635}} 
\affiliation {National Radio Astronomy Observatory, 1011 Lopezville Road, Socorro, NM 87801, USA}

\author{Amanda A. Kepley\orcidlink{0000-0002-3227-4917}} 
\affiliation{National Radio Astronomy Observatory, 520 Edgemont Road, Charlottesville, VA 22903, USA}

\author{Adam~K.~Leroy\orcidlink{0000-0002-2545-1700}}
\affiliation{Department of Astronomy, The Ohio State University, 140 West 18th Avenue, Columbus, OH 43210, USA}
\affiliation{Center for Cosmology and Astroparticle Physics (CCAPP), 191 West Woodruff Avenue, Columbus, OH 43210, USA}

\author{Sébastien Maret\orcidlink{0000-0003-1104-4554}}
\affiliation{Univ. Grenoble Alpes, CNRS, IPAG, 38000 Grenoble, France}

\author{Stuart Mumford\orcidlink{0000-0003-4217-4642}}
\affiliation{Aperio Software, Insight House, Riverside Business Park, Stoney Common Road, Stansted, UK}

\author{George C. Privon\orcidlink{0000-0003-3474-1125}}
\affiliation{National Radio Astronomy Observatory, 520 Edgemont Road, Charlottesville, VA 22903, USA}
\affiliation{Department of Astronomy, University of Florida, PO Box 112055, USA}
\affiliation{Department of Astronomy, University of Virginia, 530 McCormick Road, Charlottesville, VA 22903}

\author{\'Alvaro S\'anchez-Monge\orcidlink{0000-0002-3078-9482}}
\affiliation{Institut de Ci\`encies de l'Espai (ICE), CSIC, Campus UAB, Carrer de Can Magrans s/n, E-08193, Bellaterra, Barcelona, Spain}
\affiliation{Institut d'Estudis Espacials de Catalunya (IEEC), E-08860, Castelldefels, Barcelona, Spain}

\author{Srikrishna Sekhar\orcidlink{0000-0002-8418-9001}} 
\affiliation {National Radio Astronomy Observatory, 1011 Lopezville Road, Socorro, NM 87801, USA}

\author{Brigitta M Sipőcz\orcidlink{0000-0002-3713-6337}}
\affiliation{IPAC, California Institute of Technology, Pasadena, CA 91125, USA}

\author{A.J.M~Thomson\orcidlink{0000-0001-9472-041X}}
\affiliation{SKA Observatory, SKA-Low Science Operations Centre, 26 Dick Perry Avenue, Kensington WA 6151, Australia}

\begin{abstract}
We present the \texttt{radio-astro-tools} code suite, which consists of several Python packages that enable analysis of radio data, especially interferometric spectral cubes, in the context of the Astropy software ecosystem.
While these tools were designed with radio data in mind, they are built to be general, and have applications on data sets at other wavelengths.
The core package, \texttt{spectral-cube}, handles reading and writing and analysis of cube data, and it enables straightforward parallelization via \texttt{dask} and \texttt{joblib} backends.
Support packages include \texttt{casa-formats-io} and \texttt{radio-beam},  which handle reading of CASA tables \& images and reading and manipulation of point spread functions,  respectively.
The \texttt{pvextractor} package facilitates creation of position-velocity diagrams.
The \texttt{uvcombine} package implements the `feather' algorithm for combining single-dish and interferometric data.
The development of \texttt{radio-astro-tools} included several contributions to other repositories, including \texttt{matplotlib}, \texttt{astropy}, and \texttt{regions} to support interaction between CASA and other parts of the astronomy software ecosystem.
We also present a detailed set of tutorials written in Jupyter notebooks that can be run interactively in a browser, providing more accessible access to data analysis for radio spectral-line data cubes.
\end{abstract}

\section{Introduction}

Multiwavelength astronomy is now commonplace, especially in the current era of astronomical surveys and queue observing, in which specialization in observing in a given wavelength regime is not needed to obtain data.
Additionally, there have been significant efforts by observatories to deliver science-ready data products to astronomers, further lowering the barrier to using multiwavelength data for astronomers with a wide range of expertise.
The delivery of data from different wavelengths with different instruments but common assumptions about locations on the sky and the energy of photons drove the development of standardized world coordinate systems (WCS) \citep{Greisen2006,Calabretta2002}, providing the framework to combine these observations for analyses.
Astronomy now faces the need to handle increasingly large data volumes, and to ideally provide tools that can scale across computing infrastructures to tackle a wide range of analyses.

Data cubes are a clear example in which large data volumes push the limits of existing software and computing resources.
These cubes are now in common use among astronomers of all wavelengths, typically on a regular grid including two spatial directions and a spectral dimension (thus every spatial pixel has an associated spectrum).
In radio interferometry, the natural product of any observation is a data cube, since both spatial and spectral data are obtained simultaneously in the standard observing mode.
Radio wavelength imaging from single-dish telescopes has also naturally leaned toward cube making, often using on-the-fly (OTF) mapping  to map out large areas \citep{Mangum2007A&A...474..679M}.
In optical and infrared observing, obtaining spectra of fully sampled areas on the sky has only recently become possible, requiring either a Fourier-domain observation of a large patch of the sky with Fabry-Perot interferometers \citep[e.g.,][]{Reynolds1998} or integral field units \citep[IFUs, e.g., MUSE, SINFONI, GMOS, SAURON, OSIRIS, NIRSPEC and MIRI on JWST, SDSS-V LVM;][]{MUSEBacon2010,SINFONIEisenhauer2003,GMOSAllington-Smith2002,SAURONBacon2001, OSIRISLarkin2006,NIRSpecJakobsen2022,MIRIWright2023,LVMDrory2024,LVMKonidaris2024},
Across wavelengths, these data cubes can be large, spanning to $>100$s GB, and with ALMA's Wideband Sensitivity Upgrade \citep{Carpenter2023}, SKA pathfinders like MeerKAT, and the forthcoming ngVLA and SKA, these are expected to continue to grow by orders of magnitude in the upcoming decade.

Therefore, the widespread distribution of these imaging products creates a need for fast, memory-efficient tools to read, write, and analyze such data cubes. In particular, we need tools that allow future development to focus on methodology, improved algorithms, and techniques rather than basic file handling and low-level processes.

Here we present \texttt{radio-astro-tools}\footnote{\url{radio-astro-tools.github.io}}: a collection of Python packages built on the scientific Python ecosystem that enable memory-efficient analyses of data cubes and their associated products.
Section \S \ref{sec:packages} lists the packages that are part of the \radioastrotools ecosystem.
Section \S \ref{sec:developers} discusses how developers might use \radioastrotools.
Section \S \ref{sec:tutorials} lists the tutorials created to help users learn these tools.
And Section \S \ref{sec:integrated_ecosystem} details the key parts within the broader Astropy ecosystem that enable \radioastrotools.

\subsection{Relationship to CASA}
CASA \citep{CASATeam2022} is one of the core package for radio astronomy data reduction and analysis, having supplanted AIPS in the early 2000's for VLA data and always being the main tool for ALMA.
CASA contains many tools for cube and image analysis, which are the primary purpose of \radioastrotools that we highlight here.
While the core CASA tasks are particularly computationally efficient, the layered nature of the CASA architecture means that most calculations occur at the lower C++ layer with underlying source code that is opaque to many astronomers. In addition, the CASA's task-based Python interface means that intermediate products must be written to disk to chain multiple operations together.

With community development of standard astronomical Python packages, including \astropy \citep{ASTROPY2013,ASTROPY2018,AstropyCollaboration2022}, many astronomers now commonly use and are familiar with the scientific Python stack. 
Within the monolithic CASA distribution, CASA users face challenges with varying Python package versions and difficulty installing custom packages into CASA's Python interface. 
Recent CASA distributions in modular form, and installable via standard Python tools like pip, ease these issues and enable users to directly interface with the rest of the scientific Python ecosystem.
In many cases, \astropy has a key role in linking astronomy to other scientific packages.
\radioastrotools furthers this link to the common CASA analysis tools that radio astronomers are most familiar with.

\subsection{Relationship to astropy}
\label{sub:relation_astropy}
\radioastrotools is intended to be a radio astronomer's entry into the Astropy ecosystem.
Details of how \radioastrotools and \astropy are integrated are given in \S \ref{sec:integrated_ecosystem}.
A substantial fraction of the \radioastrotools project effort directly contributed into the core \texttt{astropy} package to provide compatibility tools or performance enhancements needed for radio astronomy work.
Some of the \radioastrotools packages are \astropy-affiliated\footnote{https://www.astropy.org/affiliated/}.
Since the inception of \radioastrotools, some of the functions built into its packages have been incorporated into \astropy core (e.g., the \texttt{NDData} system).
The long-term vision for \radioastrotools is that its code will be refactored to use more of the \texttt{NDData} system, but the interfaces provided by its packages will remain unchanged, such that a user of a package today should be able to rely on having access to the same commands and tools indefinitely.

\section{radio-astro-tools packages}
\label{sec:packages}

\radioastrotools\footnote{\url{radio-astro-tools.github.io}} is comprised of several Python packages: the main \texttt{spectral-cube} package and several light-weight packages that support specific operations or visualization for radio data.
In this section, we describe the purpose and capabilities of each package.
We provide a guide for translating between operations in CASA and \radioastrotools on our tutorial page\footnote{\url{radio-astro-tools.github.io/tutorials/}} to aid radio astronomers.

\subsection{\spectralcube}
\label{sec:spectralcube}

\spectralcube is the primary \radioastrotools package that enables fast and flexible I/O operations with most common analysis methods backed by the Astropy ecosystem.
The primary goal of \spectralcube is the seamless handling of larger-than-memory spectral data cubes, and the ability to scale processing from a user's local computer to larger jobs on computing clusters.

Users interact primarily with the \texttt{SpectralCube} class, which handles reading different data formats and types of spectral cubes.
For example, \spectralcube handles data cubes with varying spatial resolution (i.e., the beam varies between each spectral channel) as is expected for wide-bandwidth coverage and spectral scan data.

\spectralcube also offers basic support and handling for 4-dimensional full Stokes' cubes building on the \astropy stokes WCS framework.
This allows \spectralcube\ to support the gamut of polarizations used for the vast majority feed configurations offered by radio telescopes. 
This support also enables simple conversions between different polarization bases while leveraging \spectralcube's efficient memory handling.

Finally, we highlight that \spectralcube has a thorough testing suite to check all operations in the package. Users can access more information and documentation at \url{spectral-cube.readthedocs.io}.

We present an overview of how to integrate \spectralcube into new packages (\S\ref{sec:developers}), present tutorials with worked examples (\S\ref{sec:tutorials}), and describe further uses of \spectralcube with the accompanying packages described below (\S\ref{sec:integrated_ecosystem}).

\spectralcube provides easy handling for common operations and analyses of spectral data cubes. This includes the following:
\begin{itemize}
    \item Basic arithmetic between cubes with matching WCS coordinates.
    \item Handling of cube units, including intensity units that are specific to radio-astronomy, such as Janskys per beam (Jy/beam) or brightness temperature (K).  Spectral units conversion (e.g. from Hz to km/s) are also supported.
    \item Signal and noise masking. The masking framework is flexible and includes new mask creation, mask combination, and passing/combining with pre-made masks. Operations such as sigma-masking can also be used.
    \item Spatial and spectral region extraction using pixel indices, WCS coordinates, and DS9/CRTF defined regions.
    \item Moment map calculation (zeroth, first, second, Nth moments) and other common projection methods including the maximum, minimum, and spectral value at max/min brightness (e.g., velocity at peak temperature).
    \item Spectra and projection visualization.
    \item Spatial convolution to a set Gaussian beam size (using \radiobeam; \S\ref{sub:radio-beam}), median smoothing and smoothing to an arbitrary kernel type using \texttt{astropy.convolution}.
    \item Spectral smoothing with \texttt{astropy.convolution} kernels and resampling to a new spectral axis.
    \item Spatial and spectral reprojection to a given FITS header using the \reproject package.
    \item Support for full-Stokes spectral cubes.
    \item Integration with visualization packages that include glue\footnote{\url{glueviz.org/}} \citep{Robitaille2017}, yt\footnote{\url{yt-project.org/}} \citep[][]{Turk2011} and DS9\footnote{\url{ds9.si.edu}} \citep[][]{SmithsonianAstrophysicalObservatory2000}, with future plans to integrate with CARTA\footnote{\url{cartavis.org/}} \citep{Comrie2021}.
    
\end{itemize}
These operations are implemented to read in and compute the data cube in chunks, thereby avoiding reading the entire data cube when handling large data.
For operations where this is not easily done, users are warned of operations that could exceed memory limits and given the option to disable this behavior.

\subsection{\radiobeam}
\label{sub:radio-beam}

Interferometric data are generally imaged using a variant of the CLEAN algorithm \citep{Hogbom1974A&AS...15..417H}, which produces final images consisting of a model convolved with a synthesized point spread function, generally called the `clean beam' or `synthesized beam'.
\radiobeam\footnote{\url{github.com/radio-astro-tools/radio-beam},
\url{radio-beam.readthedocs.io}} is a toolkit for working with standard Gaussian clean beams.
It can read and write both single beams and per-channel beams for cubes, and includes operations for convolution, deconvolution and unit conversions (e.g., Jy/beam to K).
It also includes a general algorithm for determining the smallest common beam for convolution of a data cube to a common resolution by solving for the minimum enclosing ellipse given a set of ellipses.
Appendix \ref{app:commonbeam} provides further details on the algorithm.

\begin{center}
\begin{figure}[ht!]
    \centering
    \includegraphics{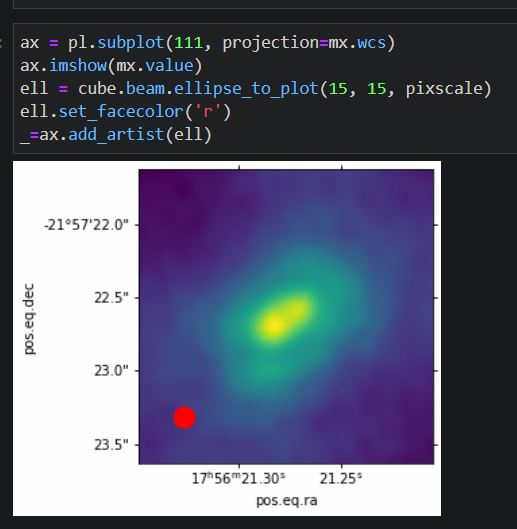}
    \caption{Example of plotting a beam with \radiobeam.}
    \label{fig:beam_plot}
\end{figure}
\end{center}

\radiobeam also includes helpful visualization routines, including plotting the beam shape on to matplotlib figures (see Figure \ref{fig:beam_plot}).

\subsection{\casaformatsio}
\label{sub:casa-formats-io}

CASA natively writes data into directories that contain tabular data, and along with metadata, this format is used for both data cubes, images, and visibility data (typically referred to as measurement sets in CASA).
The CASA format has historically only been readable by \texttt{casacore}\footnote{\url{casacore.github.io/casacore/}} \citep{Casacore}, which is a library written in C with bindings available in several other languages.
\texttt{casacore} does many things besides file reading, though, and can be challenging to install\footnote{\url{newton.cx/~peter/howto/access-casa-in-python-without-casapy/}}.

To overcome this limitation and provide an interface that optimally leverages other packages in the Python ecosystem, we provide \texttt{casa-formats-io}\footnote{\url{casa-formats-io.readthedocs.io}, \url{github.com/radio-astro-tools/casa-formats-io}}, a table reader for CASA data formats written in Python and C (though all user-facing functions are in Python).
\texttt{casa-formats-io} is cross-platform, operating on Unix, Windows, and macOS operating systems, and uses \texttt{dask} \citep{Dask} to lazily load the data following CASA's chunking scheme.
To our knowledge, \texttt{casa-formats-io} is the first CASA format reader that will work on Windows operating systems.

For data cubes and images, \texttt{casa-formats-io} loads CASA tables as Dask Arrays. For example, with \spectralcube, users load and manipulate data cubes with no difference in syntax, apart from specifying the CASA format for the input data upon reading the data.

For visibility data, \texttt{casa-formats-io} loads CASA tables as Astropy Tables\footnote{\url{docs.astropy.org/en/stable/table}}, providing access to an extensive library for tabular data handling.
In this way, CASA measurement sets can now natively read with a wider variety of visualization and analysis tools, including \texttt{glue}.

\subsection{\uvcombine}
\label{sub:uvcom}
Interferometric observations do not fully sample the Fourier plane of the sky; they intrinsically leave a gap at the center of the domain, which is referred to as `short spacing' or the `DC component.'
To account for missing flux from the short spacing information, observations taken with a single filled-aperture telescope (`single-dish') can be combined with interferometric images.
CASA provides a tool for the Fourier-space combination of interferometric and single-dish images in the \texttt{feather} task.
The \texttt{uvcombine}\footnote{\url{uvcombine.readthedocs.io}} package provides a Python-based implementation of the same algorithm.
Our implementation includes additional configurable options, can process a wider variety of input data types, and converts between different units in the input data.

\uvcombine also includes tools to estimate the flux scaling factors between the single-dish and interferometric data, as described in \citet{Stanimirovic2002} and \citet[][see Appendix A]{Koch2018}.
These tools compare the relative flux at scales that both data sets are sensitive to, which is necessary to test whether the flux calibrations applied are consistent or whether a correction factor should be applied (more often to the single dish data).

We test \texttt{uvcombine} against CASA's feather task and find consistent results.
Appendix \ref{app:feather_compare} shows the agreement between both methods using a synthetic fractal image, where the differences are negligible for any reasonable level of observational noise and flux scaling uncertainty.
Applications to feathering real data are consistent with this synthetic example \citep[e.g.,][]{Koch2025}.

\subsection{\pvextractor}
\label{sub:pvextractor}
Position-velocity diagrams are two-dimensional slices through three-dimensional data cubes, where the third dimension is spectral and can generally be interpreted, for a given observed emission or absorption line, as Doppler-shifted velocity of gas.
The \texttt{pvextractor} package\footnote{
\url{github.com/radio-astro-tools/pvextractor/},
\url{pvextractor.readthedocs.io}}
allows users to produce slices along arbitrary paths.
It also allows averaging perpendicular to the selected paths.
Slice paths can be specified in pixel or world coordinates and can be produced either programmatically or through the \texttt{PVSlicer} GUI.
\texttt{pvextractor} is incorporated into the \texttt{glue} package.

\subsection{Additional packages and functionality from \radioastrotools}

\subsubsection{statcont}

\texttt{statcont} was developed to enable estimation of the continuum flux from complex ``line-forest'' data sets and is described in full in \citet{Sanchez-Monge2018}.
The original package used only \texttt{astropy.io.fits} and required loading the entire data set into memory.
We have added a capability to use \texttt{statcont} with \texttt{spectral-cube}, which enables running \texttt{statcont} on cubes that are larger than the computer's memory.
We note that a key part of this process was significant performance enhancements to the \texttt{astropy.stats.sigma\_clip} implementation.

\subsubsection{Interactive drawing in astropy regions}
\label{subsub:regions}
The \texttt{regions} package\footnote{\url{astropy-regions.readthedocs.io/en/stable/}, \url{github.com/astropy/regions}} is an Astropy coordinated package and is not direct part of \radioastrotools, however, it contains some functionality that was developed alongside \radioastrotools\ and is familiar to CASA users.

\texttt{regions} supports reading and writing the CRTF file format (CASA Region Text Format) which is used internally by CASA to specify regions of interest.
This enables reading in regions created in the CASA viewer or CARTA that similarly support CRTF \citep{Comrie2021}, and conversions to the also-popular DS9 region format.

With interactive region drawing supported via \texttt{matplotlib} \citep{Matplotlib}, \texttt{regions} supports interactive drawing and manipulation of regions within a Jupyter notebook or IPython terminal, where the region can be exported with the corresponding WCS information.
One major use case for this functionality is the ability to interactively draw and save region files that can be reproducibly used as clean masks, for example with CASA's \texttt{iclean} and \texttt{tclean} tasks.


\begin{center}
\begin{figure}[t!]
    \centering
    \includegraphics[width=0.9\textwidth]{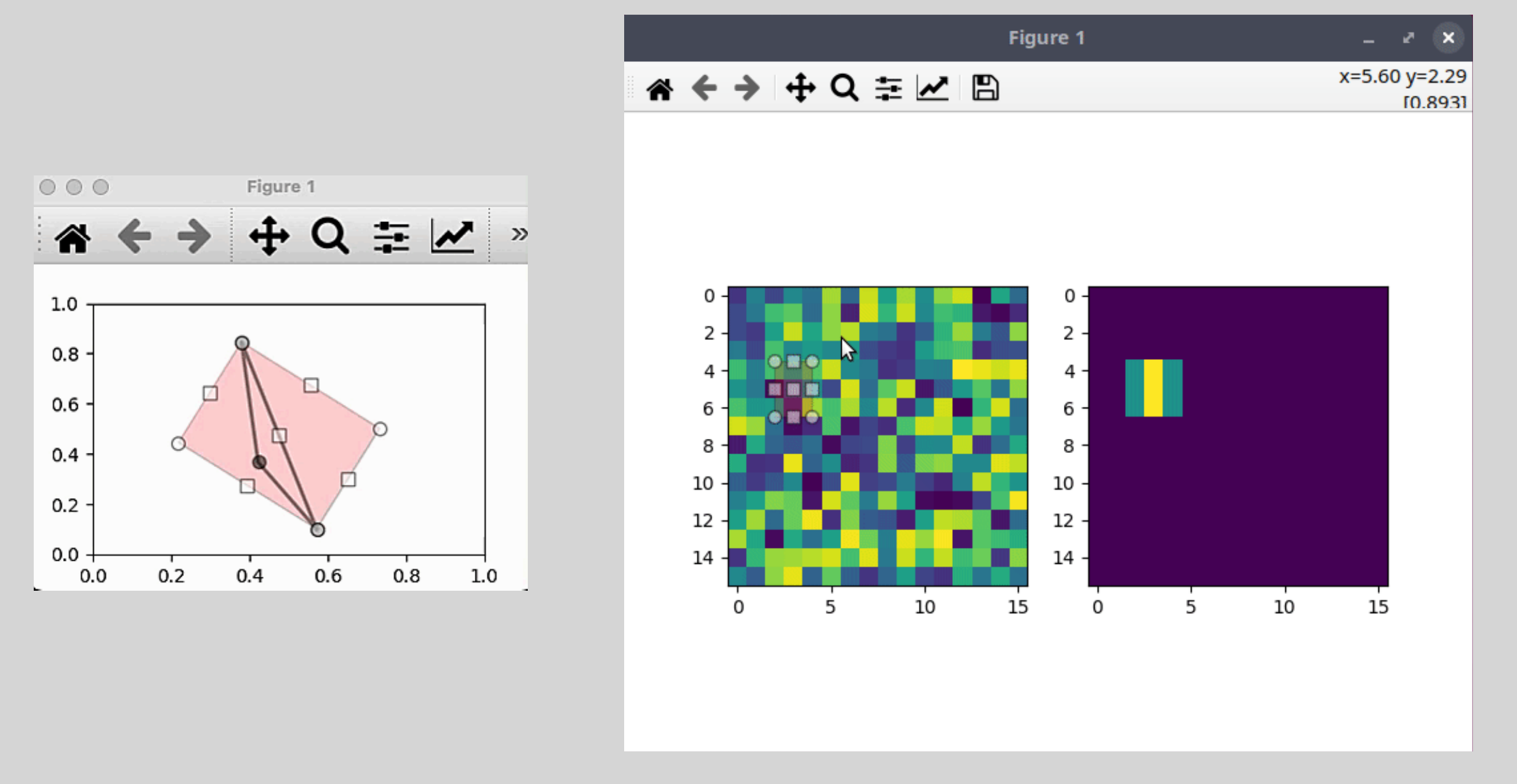}
    \caption{Static images of interactive region drawing with \matplotlib and \texttt{regions}.
    Left: Square region on a blank canvas (\href{user-images.githubusercontent.com/6197628/148653577-57ef2750-09dc-4a13-bcd7-f2c23a58e3b7.gif}{Interactive Version}).
    Right: Square region projected onto an image (\href{user-images.githubusercontent.com/314716/82560196-dd4df800-9b68-11ea-8ff3-08dc2fc4ac14.gif}{Interactive Version}).
    Examples from D. Homeier and D. Stansby.}
    \vspace{0.75cm}
    \label{fig:region_examples}
\end{figure}
\end{center}


\section{Using radio-astro-tools for development of other packages}
\label{sec:developers}

The \radioastrotools packages are ideally suited to handle I/O operations in other Python packages.
This is a significant strength given the efficient larger-than-memory data handling in \spectralcube and parallelization of operations with the \dask integration that users can build upon.
We include an example use case in our set of tutorials demonstrating how \spectralcube and \dask can parallelize fitting a Gaussian model to every spectrum in an ALMA data cube, loading only small chunks into memory at a time (\S\ref{sec:tutorials}).
This example overcomes an often-faced issue for some spectral cube modeling tools, which converted standard cube file formats (FITS, CASA image) to non-standard forms (pickle or text file) to boost performance.
\spectralcube removes the need for these extra conversion and I/O steps to allow researchers to focus on algorithmic innovations related to the science or analysis goal.

\begin{figure}
    \centering
    \includegraphics[width=0.5\textwidth]{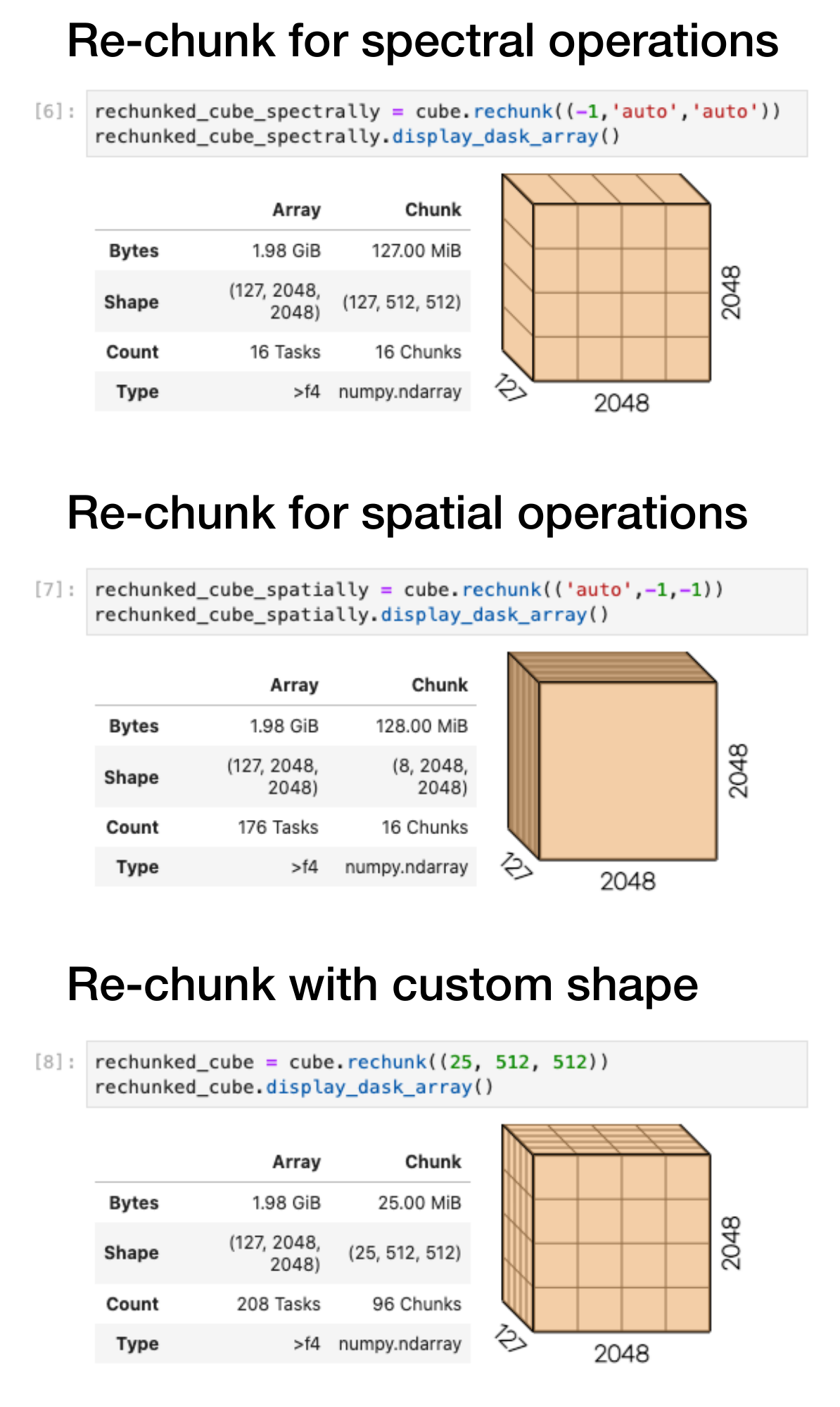}
    \caption{Examples of rechunking a data cube to optimize for spectral and spatial operations (top, middle), and passing a custom chunk shape (bottom) using the \dask integration in \spectralcube.
    }
    \label{fig:dask_rechunks}
    \vspace{0.5cm}
\end{figure}

Figure \ref{fig:dask_rechunks} shows a visual example of spectral cubes using the \texttt{dask} backend. \texttt{dask} provides the ability to define differently shaped chunks that, depending on the analysis operation, may be more efficient when loading in chunks of the data.
For example, operations on individual spectra (e.g., spectral smoothing, regridding, and fitting) are the most efficient loading in individual or groups of spectra.
In contrast, operations on the spatial axes (e.g., spatial convolution) are most efficient loading in individual channels in the data cube.
\spectralcube provides the necessary links for users to quickly rechunk their data, including using intermediate temporary files with \texttt{zarr}\footnote{\url{zarr.dev}}, and efficiently perform a wide variety of analyses.
Our tutorials on reprojecting data cubes provide an example of this functionality (\S\ref{sec:tutorials}).

The \spectralcube documentation\footnote{\url{spectral-cube.readthedocs.io/en/latest/developing_with_spectralcube.html}} includes further information and recommended practices for incorporating \radioastrotools for development in other packages.

We note that \radioastrotools is already used in a number of calibration, imaging and analysis pipelines.
Examples include ALMA \citep[e.g., ALMA-IMF\footnote{\url{github.com/ALMA-IMF/reduction}}; PHANGS\footnote{\url{github.com/akleroy/phangs_imaging_scripts}}; ALMA-JELLY; ACES\footnote{\url{https://github.com/ACES-CMZ}};  COMPASS\footnote{\url{https://github.com/compass-alma-large-program/compass-data-notebooks}}; sSCALES;][Leroy et al. in prep;]{Ginsburg2021,Leroy2021, Souchereau2025,Ginsburg2026,Jorgensen2026}, 
GBT \citep[GAS\footnote{\url{github.com/GBTAmmoniaSurvey/GAS}};  DEGAS\footnote{\url{github.com/GBTSpectroscopy/degas}}][, Kepley et al. in prep.]{Friesen2017}, ASKAP \citep[e.g., GASKAP; ][]{Pingel2021} and VLA \citep[e.g., LGLBS \footnote{\url{github.com/LocalGroup-VLALegacy}}][]{Koch2025} large programs ,
and the calibration pipeline for Apertif \citep[Apercal; ][]{Adebahr2022}.

\subsection{glue}
The \texttt{glue} visualization and analysis toolkit uses many components of the \texttt{radio-astro-tools} package both for reading files and for analysis.
For example, the position-velocity extraction in \texttt{glue} directly uses \texttt{pvextractor}.
Using \casaformatsio, we have demonstrated that CASA tables can be loaded into \texttt{glue}, enabling direct analysis of measurement sets.
This functionality is available in the development version, but this remains experimental.

\section{Tutorials}
\label{sec:tutorials}

We have written a series of tutorials with fully-worked analyses for common operations of spectral-line data cubes.
The tutorials are written as Jupyter notebooks and can be accessed and run interactively in a web browser using Binder\footnote{\url{mybinder.org}} with data hosted on Zenodo\footnote{\url{zenodo.org}} and the ALMA data archive.
This completely removes the need to both install various packages and local downloading of data.
Additionally, two of our tutorials are incorporated into the astropy learn project\footnote{\url{learn.astropy.org}}.
These tutorials are written at a suitable level for senior undergraduate or graduate students, or astronomers seeking to become more familiar with analyzing radio astronomy data.
The tutorial set is intended to provide a common source on often-used analysis routines that are either available in other software without a pedagological example, or are commonly re-invented based on literature sources (e.g.,  cases where graduate students often re-write the same algorithms for a common purpose).

The tutorial site\footnote{\url{radio-astro-tools.github.io/tutorials}} is hosted on GitHub and hosts rendered versions of the tutorials, download links to the notebooks on GitHub, and direct links to open interactive sessions of the tutorials on Binder.
The current tutorial set includes:
\begin{enumerate}
    \item Fitting spectral line and spatial models using \spectralcube and \texttt{astropy.modeling}.
    \item A guide to \spectralcube for CASA users.
    \item Signal-masking and calculating moment maps from spectral cubes.
    \item Parallel fitting of spectral line models with \spectralcube, \astropy and \texttt{dask}.
    \item Creating and plotting Position-velocity (PV) diagrams.
    \item Identification of the far 3 kpc arm in the \citet{Dame2001ApJ...547..792D} all-sky CO map, reproducing the results from \citet{Dame2008ApJ...683L.143D}.
    \item Position-velocity (PV) diagram of a protoplanetary disk using public data from the MAPS ALMA large program\footnote{\url{alma-maps.info}} \citep{Oberg2021ApJS..257....1O}.
    \item Reprojecting two data cubes to a common resolution, projection and pixel grid.
    \item Reprojecting two data cubes and applying a signal mask from one cube to another.

\end{enumerate}

We plan on expanding this suite of tutorials and welcome community-contributed tutorials, which we strongly encourage to be submitted and incorporated into the astropy learn project.

\begin{figure}
    \centering
    \includegraphics[width=\textwidth]{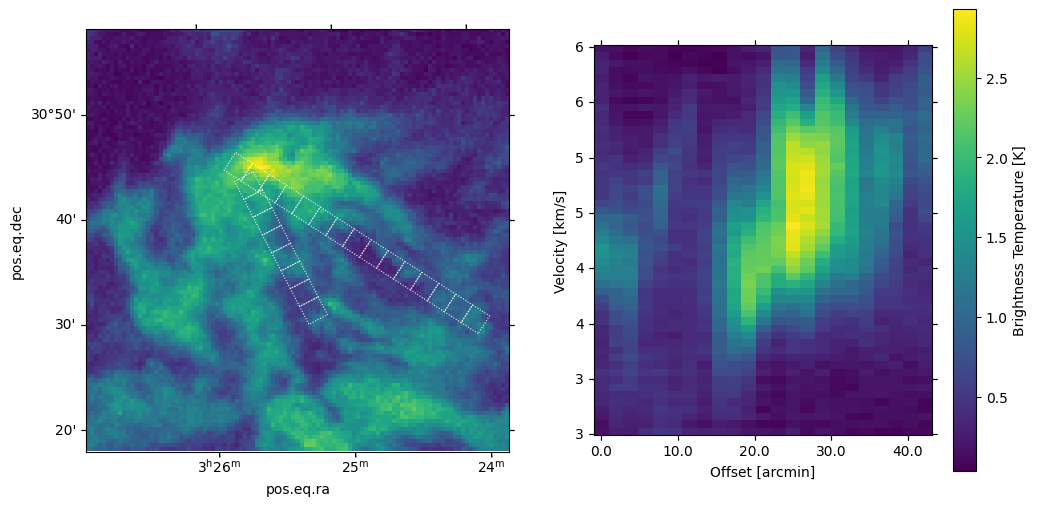}
    \caption{Example position-velocity diagram (right) and trace (left) produced in our astropy learn tutorial.
     The white rectangles shown
     in the left figure are extracted from the data cube and spatially averaged.
     The average spectrum is shown in intensity as a vertical slice in each column of the 
     right panel.
    }
    \label{fig:pvexample}
    \vspace{0.5cm}
\end{figure}

\section{The integrated ecosystem for radio astronomy}
\label{sec:integrated_ecosystem}

The \radioastrotools project aims to integrate the radio astronomy analysis tools with the rest of the astronomical Python ecosystem.
To achieve this goal, we supply the tools for reading, manipulating, and writing astronomical images, cubes, spectra, and regions of interest from a wide variety of data sources.
In this section, we summarize important links between the astronomical Python ecosystem and their integration with \radioastrotools packages.

\subsection{Reprojecting images with \reproject}

Pixel-by-pixel comparison of data from different observations with varying pixel scales is often needed.\footnote{\url{github.com/radio-astro-tools/tutorials/blob/master/SpectralCubeReprojectExample.ipynb}} 
The cube reprojection tutorial (\S\ref{sec:tutorials}) demonstrates how users can smooth and regrid two archival ALMA data cubes onto a common spatial and spectral grid.
All aspects of the smoothing and regridding, both spatial and spectral, can be handled within \spectralcube using \texttt{astropy.units} for unit handling, \texttt{astropy.convolution} for convolution, and \reproject\footnote{\url{reproject.readthedocs.io/}} for regridding onto new world coordinate systems.
\reproject can also handle Dask Arrays, providing a way to parallelize and efficiently manage the reprojection of large data cubes.

\subsection{Regions-of-interest with \regions and interactive \matplotlib tools}

Selecting portions of the sky, or simply drawing on the sky, is an integral component of data analysis and publication figures.
There are several region file formats and editors of varying popularity, particularly the SAOImage DS9 visualization tool \citep{SmithsonianAstrophysicalObservatory2000}, which defines a \texttt{.reg} format, and the (now deprecated) CASA Viewer and CARTA visualization tools \citep{Comrie2021}, which primarily use the CRTF region format\footnote{\url{casadocs.readthedocs.io/en/stable/notebooks/image_analysis.html}, \url{casaguides.nrao.edu/index.php/CASA_Region_Format}}.
In the future, we intend to support the proposed general region specification format defined by the International Virtual Observatory Alliance (IVOA), though we are not aware of any current code implemented to use it.
The astropy \regions package\footnote{\url{github.com/astropy/regions/}} implements a unified interface to CRTF and DS9 regions, and can read and write both formats.

\spectralcube uses the \regions functionality for several operations, including the ability to mask out and return spatial subsets of the data only within the region of interest.
Uses of this region-based masking including averaging spectra over a source of interest and creating position-velocity diagrams across irregular shapes (demonstrated in the PV diagram tutorial; \S\ref{sec:tutorials}).

Regions can also be used in interactive cleaning to specify where model components should be added or disallowed.
On-going development will complete the \texttt{matplotlib} functionality to interactively define sky regions for use with deconvolution, providing useful functionality that can be incorporated into new CASA development of an interactive clean GUI\footnote{\url{github.com/casangi/casagui}}, which replaces similar functionality in the now deprecated casaviewer.
New additions to \texttt{matplotlib} allow the creation and editing of region files interactively with the matplotlib viewer, which is agnostic of backend - i.e., it can be used on any matplotlib-compatible interactive viewer, including notebook viewers.
The regions created in the interactive viewer can be saved in world coordinates when Astropy's \texttt{wcscaxes} is used to display the regions.

\subsection{All CASA-produced images are spectral cubes}

CASA writes all of its outputs (its \texttt{.image}, \texttt{.residual}, etc.) files as 4-dimensional cubes:
the first two dimensions represent sky coordinates, the third spectral coordinates, and the fourth the Stokes axis for polarization.
For many images, i.e., continuum images or single-polarization Stokes I cubes, there are one or more `degenerate' axes,
i.e., axes with length 1.  CASA explicitly acknowledges this by including the \texttt{dropdeg} keyword
in the \texttt{exportfits} function, which will output a FITS file with no header information for the axes with length
1. However, for any data files not exported in this way, or any native CASA files, the underlying object is still 4-dimensional.
That means that these objects can be opened directly with spectral cube, even if they are just 2-dimensional images.

\subsection{Relation to other Python packages handling spectral-line data cubes}

Multiple Python packages support spectral-line data cubes. \spectralcube and the broader \radioastrotools package suite provide general handling for cube data, but are specifically aimed at the radio, mm, and sub-mm wavelength regime. However, data formats used in other wavelength regimes (particularly IFU data cubes) are readily handled with \spectralcube.

We note here the relationship between \spectralcube and two commonly-used spectral-line data packages in the Astropy ecosystem: \texttt{specutils} \citep{Astropy-SpecutilsDevelopmentTeam2019} and \texttt{pyspeckit} \citep{pyspeckitGinsburg2022}.

\texttt{specutils} \citep{Astropy-SpecutilsDevelopmentTeam2019} includes functionality first implemented in \spectralcube. Though currently independent, the goals and development remain closely aligned with \spectralcube\ with \texttt{specutils} generally focused on shorter wavelength regimes and reading and writing lower-dimensional spectral formats (e.g., 1D and multi-object spectra), though it can also load cubes.
The main difference in cube handling between \texttt{specutils} and \spectralcube is that \spectralcube focuses on handling large cubes, while \texttt{specutils} does not make any special provisions for file size.
As noted in \S\ref{sub:relation_astropy}, the long-term goal of both packages is to incorporate a shared backend based on NDDdata in \texttt{astropy}.

\texttt{pyspeckit} is primarily focused on spectral modeling of spectral data and can load data from \spectralcube.
Though fitting of spectral models can be done with \spectralcube objects (\S\ref{sec:tutorials}), fitting is not a component of the \radioastrotools\ suite.
\texttt{pyspeckit} includes a suite of preconfigured models and a GUI system for running those fits, which is not a component of the \radioastrotools suite.
Thus, \texttt{pyspeckit} should be thought of as leveraging but extending upon \spectralcube's capabilities.

\section{Summary}

The \texttt{radio-astro-tools} project provides a wide range of tools for spectral cube analysis, and integrates with the broader astronomical Python ecosystem (e.g., Astropy) to enable analysis of multiwavelength data in a common framework. This paper describes the range of packages, common use cases, and a set of tutorials to enable broader access to radio interferometric data products.
By providing this common framework, we aim to allow astronomers to develop new and improved algorithms for the increasingly-complex data cubes that modern interferometers can provide, rather than being forced to individually tackle common file and memory-handling operations that \texttt{radio-astro-tools} can handle.

The \radioastrotools project continues to be developed and we particularly welcome community contributions and input on the continued development.
Our current development also includes expanding functionality, including handling for full polarization products that is already being used in tools like \texttt{plumber}\footnote{\url{github.com/ARDG-NRAO/plumber}} to generate full Stokes primary beam models for radio interferometers.
Additionally, our on-going development will allow for interactive editing tools with the \texttt{regions} packages and its incorporation into the \texttt{glue} visualization and analysis package.

Our future development plans include additional integration with the Radio Astronomy Data Processing System (RADPS), which is a program initiated by NRAO to develop a system for processing data from the ALMA Wideband Sensitivity Upgrade \citep[WSU;][]{Carpenter2023} and ngVLA \citep{Murphy2018}. RADPS will replace the functionality of the current CASA and NRAO Pipelines.
The RADPS prototype\footnote{\url{github.com/casangi}} uses \texttt{xarray}\footnote{\url{xarray.dev}} that is also based on \texttt{dask} and the use of \texttt{casa-formats-io} is being investigated to handle CASA legacy table and image formats.

\begin{acknowledgments}
We gratefully acknowledge all community contributions to \radioastrotools and the support from two ALMA development studies critical to the development of \radioastrotools: ``Community Science Tool Development'' (Cycle 2; PI: A. Leroy) and ``Linking CASA to the Astropy ecosystem'' (Cycle 5; PI: A. Ginsburg).
We also acknowledge discussions with all participants of the ``Spatio-spectral Modeling of Interferometric Data: Preparing for the Wideband Era'' workshop, which was partially supported by the North American ALMA Science Center (NAASC). The National Radio Astronomy Observatory and Green Bank Observatory are facilities of the U.S. National
Science Foundation operated under cooperative agreement by Associated Universities, Inc.
A.S.-M.\ acknowledges support from the RyC2021-032892-I grant funded by MCIN/AEI/10.13039/501100011033 and by the European Union `Next GenerationEU’/PRTR, as well as the program Unidad de Excelencia María de Maeztu CEX2020-001058-M, and support from the PID2023-146675NB-I00 (MCI-AEI-FEDER, UE).
A.G. acknowledges support from the NSF under grants AAG 2008101 and 2206511 and CAREER 2142300, and from the National Aeronautics and Space
Administration (NASA) under the ROSES-2020 OSTFL program and ROSES-24 HPOSS program, grant nos. 80NSSC22K0347 and 80NSSC25M7029, to the astropy project.
J.D.H acknowledges funding from the European Research Council (ERC) under the Horizon Europe framework programme. This work is supported by an ERC Advanced Grant (Acronym: eCMZs, Number 101201230). Funded by the European Union. Views and opinions expressed are, however, those of the author(s) only and do not necessarily reflect those of the European Union or the European Research Council Executive Agency. Neither the European Union nor the granting authority can be held responsible for them. 

\vspace{0.2cm}
\textit{Software: } 
aplpy \citep{APLpyASCL,APLpyZenodo}, 
astropy \citep{ASTROPY2013,ASTROPY2018,AstropyCollaboration2022},
CASA \citep{CASATeam2022}, 
dask \citep{Dask},
ds9 \citep{SmithsonianAstrophysicalObservatory2000},
glue \citep{Glue}, 
fsspec \citep{fsspec}, 
joblib \citep{joblib},
matplotlib \citep{Matplotlib}, 
numpy \citep{NumPy}, 
reproject \citep{Reproject}, 
regions \citep{Regions},
scikit-image \citep{scikitimage}, 
scipy \citep{SciPy}, 
tqdm \citep{tqdm}, 
yt \citep{Turk2011}, 
 zarr \citep{Zarr}
 
\end{acknowledgments}

\appendix

\section{Approximations in solving for the common beam in \texttt{radio-beam}}
\label{app:commonbeam}

A common operation for radio data is to convolve to a common resolution, either internal to the data (e.g., between spectral channels) or between different data sets.
\radiobeam includes two algorithms to solve for the minimum-enclosing common beam: (i) an exact solution for sets of two beams following the implementation in \texttt{casacore}, and an approximation for more than two beams based on the Khachiyan algorithm \citep{Khachiyan1993}.
Here we describe our implementation and the approximations made to enable fast computation for large ($>1000$) sets of beams.
Further information is given in the documentation\footnote{\url{radio-beam.readthedocs.io/en/latest/commonbeam.html}}.

Finding the common beam is equivalent to solving for the minimum enclosing ellipse given a set of ellipses.
The formal $N$d ellipsoid solution to this problem requires convex optimization as the solution lies at the edge of the valid parameter space \citep[``minimally'' enclosing;][ Section\footnote{\url{http://web.stanford.edu/~boyd/cvxbook/}} 8.4]{boyd2004convex}.
Our problem for the common beam is simplified in that it is only in 2D, all ellipses are centered at the origin, and we require only loose convergence to much less than pixel size since most data sample $\sim3-7$ pixels per full-width-half-max (FWHM).
We found that the Khachiyan algorithm \citep{Khachiyan1993,todd2007khachiyan} was well-suited given these simplifications.
The Khachiyan algorithm samples points along the edge of each ellipse in the set.
From these points, we compute the convex hull of the set to define the boundaries and use those boundary points to solve for the minimum enclosing ellipse.

\radiobeam checks that all beams in the set can be deconvolved by the common beam solution.
However, the algorithm can converge within the allowed tolerance to marginally smaller than the true enclosing ellipse.
To avoid these cases where a beam cannot be marginally deconvolved, we increase the boundaries used by the Khachiyan algorithm by a relative fraction $1+\epsilon$ ($\epsilon=0.001$ by default).
$\epsilon$ is allowed to incrementally increase to a set maximum (default of $0.01$) to ensure convergence to a common beam that can deconvolved from each beam in the set.
While the addition of $\epsilon$ increases the common beam area, the default limits ensure that the increase should be far smaller than the typical pixel size and therefore negligible.

The time to compute the common beam with our implementation, including consistency checks, is far shorter than most cube operations.
With a set of 1178 beams, the computation takes $0.7$~s with the default parameters.

\section{Feathering comparison between \texttt{UVCOMBINE} and CASA}
\label{app:feather_compare}

\begin{figure}
    \centering
    \includegraphics[width=\textwidth]{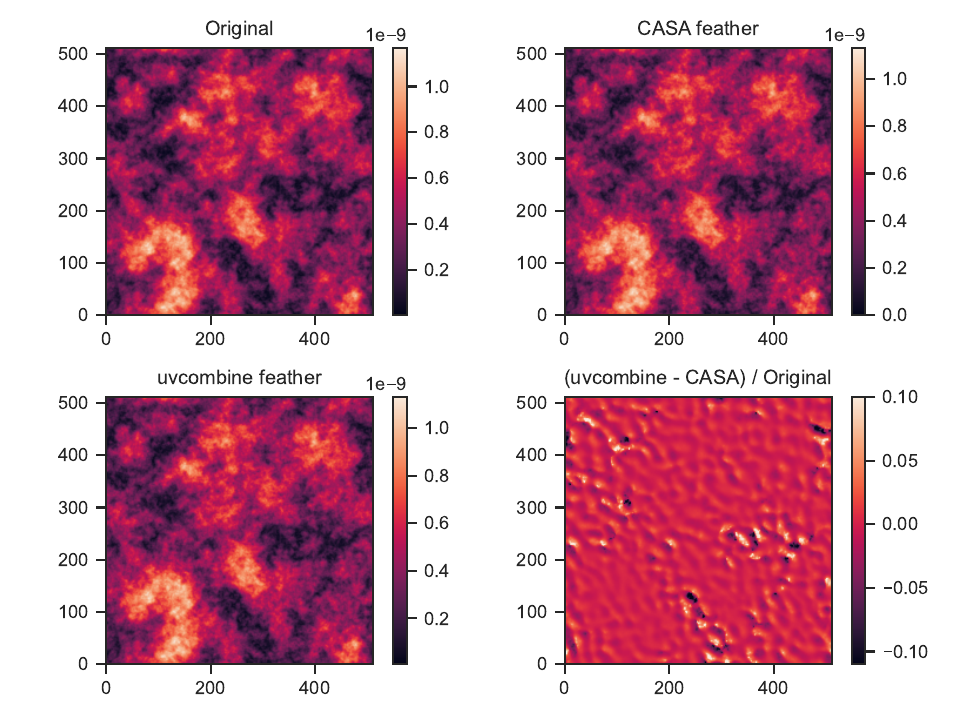}
    \caption{Demonstration of equivalent feathered maps using \uvcombine and CASA on a generated power law data.
    Only small deviations remain between the feathered maps relative to the original image.
    }
    \label{fig:feathercompare}
\end{figure}

We demonstrate that \uvcombine and CASA's feather task produce equivalent combined maps by generating an image with a known power law and simulating a single-dish and interferometer response.
The top left panel in Figure \ref{fig:feathercompare} shows the original image with a power law index of $-3$.
We simulate a single-dish observation by convolving the data to a beam size of $15''$ (using a Gaussian kernel).
For the interferometer response, we assume an idealized case of a filled uv-plane between scales of $2\mbox{--}40''$, or equivalently a low- and high-pass top-hat filter.
We use the same inputs when feathering: a scale factor of $1.0$ and no low pass filtering of the single-dish data on small scales.
The feathered data, shown in the top right and bottom left in Figure \ref{fig:feathercompare}, are visually identical.
The bottom right panel shows the difference between the feathered data, normalized by the original image.
The feather maps are consistent where the simulated emission is brightest, and small deviations persist only for the faintest features.
We find that the power law indices of the feathered maps are consistent within $<1\sigma$ uncertainty.


The continuous integration testing for \uvcombine includes tests comparing to the output from CASA's feather task on simulated power law, similar to this example.

\bibliography{spectral-cube}{}

\end{document}